\documentclass{IEEEcsmag}
\pdfoutput=1
\usepackage[colorlinks,urlcolor=black,linkcolor=blue,citecolor=blue]{hyperref}
\usepackage{amsthm}
\usepackage{lipsum}
\usepackage{listings}
\usepackage{siunitx}
\usepackage{upmath}
\usepackage{xifthen}
\usepackage{xspace}

\newtheorem{definition}{Definition}

\newcommand{\TODO}[1]{{\color{magenta} \textbf{TODO}\ifthenelse{\equal{#1}{}}{\xspace}{:~}#1 }}

\newcommand{\URL}[1]{\href{https://#1}{#1}}
\newcommand{\DEBBUG}[1]{\href{https://bugs.debian.org/cgi-bin/bugreport.cgi?bug=#1}{bugs.debian.org/#1}}
\newcommand{\BUILDINFO}{\texttt{.buildinfo}\xspace}

 \jvol{XX}
\jnum{XX}
\paper{8}
\jmonth{May/June}
\jname{IEEE Software}
\pubyear{2021}

\begin{document}

\title{Reproducible Builds: Increasing the Integrity of Software Supply Chains}
\author{Chris Lamb}
\affil{Reproducible Builds}

\author{Stefano Zacchiroli}
\affil{Université de Paris and Inria, France}

 \begin{abstract}
  Although it is possible to increase confidence in Free and Open Source
  Software (FOSS) by reviewing its source code, trusting code is not the same
  as trusting its executable counterparts. These are typically built and
  distributed by third-party vendors, with severe security consequences if
  their supply chains are compromised.
In this paper, we present \emph{reproducible builds}, an approach that can
  determine whether generated binaries correspond with their original source
  code. We first define the problem, and then provide insight into the
  challenges of making real-world software build in a ``reproducible''
  manner---this is, when every build generates bit-for-bit identical results.
Through the experience of the \emph{Reproducible Builds project} making
  the Debian Linux distribution reproducible, we also describe the affinity
  between reproducibility and quality assurance (QA).
\end{abstract}

 \maketitle

\label{sec:intro}

\begin{quote}\itshape
You can't trust code that you did not totally create
  yourself. [\ldots] No amount of source-level verification or scrutiny will
  protect you from using untrusted code.
\begin{flushright}\normalshape
    --- Ken Thompson (1984)
\end{flushright}
\end{quote}

\bigskip

\chapterinitial{How can we be sure} that our software is doing only what it is
supposed to do? This was the key takeaway from Ken Thompson's 1984 Turing
Lecture, ``Reflections on Trusting Trust''~\cite{thompson-trust}. But with
people today executing far more software than they compile, the number of users
who ``totally create'' software they run has dropped dramatically since then.

Let us narrow the issue to Free and Open Source Software (FOSS), where all
source code is freely available. Hypothetically, users can examine the source
of all the software they wish to use in order to confirm it does not contain
spyware or backdoors---indeed, one of the original promises of FOSS was that
distributed peer review~\cite{storey2012-peerreview} would result in enhanced
end-user security. However, whilst users \emph{can} inspect source code for malicious
flaws, almost all software is now distributed as pre-built binaries.
This permits nefarious actors to compromise end-user systems by
modifying ostensibly secure code during its compilation or distribution.

For example, a Linux distribution might compile ``safe'' software on
compromised servers and unwittingly spread malicious executables onto countless
systems. Other vectors include engineers being explicitly coerced
into incorporating vulnerabilities, as well as the covert compromise of
developers' computers (remotely or through ``evil maid''
attacks~\cite{rutkowska2009evil}) so they unwittingly distribute tainted
binaries via app stores and other channels.

Software supply-chain attacks are no longer hypothetical scenarios. In December
2020, news broke
that attackers subverted the
SolarWinds Orion software to inject malicious code into executables at
build time, resulting in a severe data breach across several US government
branches~\cite{zdnet2021solarwinds}. 174 similar attacks have been detailed
in the literature too~\cite{ohm2020osssupplychain}. Due to their potential
impact, software supply chains have become a high-value target in recent years,
and this trend appears to be accelerating.

Practical and scalable solutions to these attacks are therefore urgently
needed, and an approach known as \emph{reproducible builds} is one such
countermeasure. However, it is only applicable if the software sources are
widely available---although malware can be detected directly in
binaries~\cite{ye2017malwaresurvey, amoroso2018secprogress}, doing so is
inefficient when source code is available for audit.

The key idea behind the reproducible builds (\emph{R-B}) approach is that, if
we can guarantee that building a given source tree always generates bit-for-bit
identical results, we can establish trust in these artifacts by comparing
outputs acquired from multiple, independent builders.

In this paper, we present the R-B approach from the perspective of
software professionals. We show how software \emph{users} can benefit from the
increased trust in executables they run as well as how \emph{developers} and
build engineers can help make software reproducible. We also describe the
\emph{quality assurance} (QA) tools available to improve build reproducibility,
highlighting how they further mutually-beneficial goals such as reducing build
and test ``flakiness''~\cite{durieux2020flakybuilds, luo2014flakytests}. This
paper is informed in large part by the experience of the Reproducible Builds
project (\URL{reproducible-builds.org}), a non-profit initiative that
popularized the R-B approach.

 \section{REPRODUCIBLE BUILDS}
\label{sec:trust}

The core element of the reproducible builds model is the following property:

\begin{definition}
  \label{def:r-b}
  The build process of a software product is \emph{reproducible} if,
  after designating a specific version of its source code and all of its build
  dependencies, every build produces \emph{bit-for-bit identical artifacts}, no
  matter the environment in which the build is performed.
\end{definition}

\begin{figure*}
  \centering
\includegraphics[width=\textwidth]{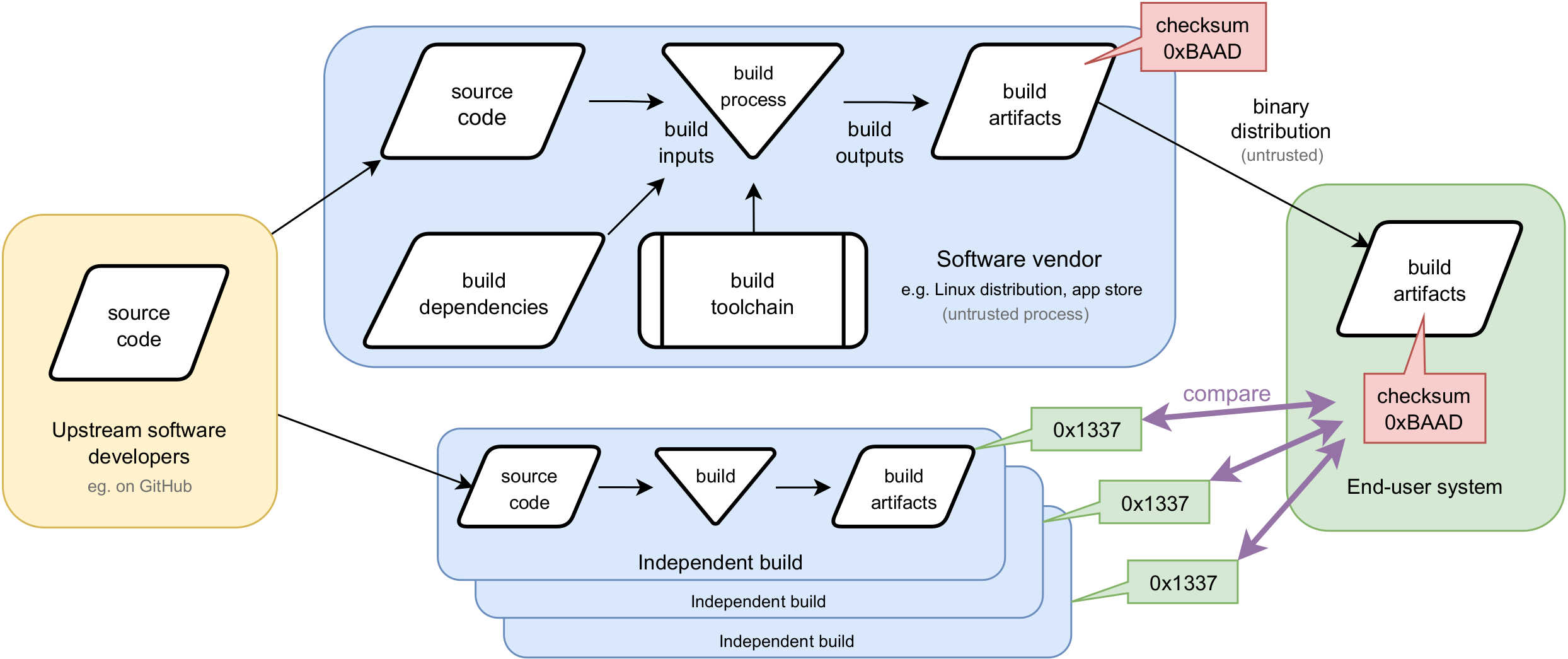}
  \caption{The \emph{reproducible builds} approach to increasing trust in executables
    built by untrusted third parties. The end-user should reject the binary
    artifact from their software vendor, as its checksum (\texttt{0xBAAD})
    does not match the one built by multiple, independent third-parties (\texttt{0x1337}).}
  \label{fig:approach}
\end{figure*}

In other words, once we reach an agreement on the exact software version(s) we
wish to build, anyone who builds that software should always generate precisely
the same artifacts.

Figure~\ref{fig:approach} shows how users can leverage this property to
establish trust in FOSS executables. Software development happens upstream as usual
(e.g.~on platforms such as GitHub and GitLab) and, from there, source code
reaches downstream vendors such as Linux distributions and app stores. These vendors then
build binaries from these sources, which are subsequently distributed to
end-users. Note that neither the distribution nor the build process are
completely trusted in this scenario, reflecting the hostile environment of the
real world.

When software builds reproducibly, however, we can still establish trust in these
executables. This is because users can corroborate whether their
newly-downloaded binaries are identical to those that \emph{others} have built
themselves.

How this works is as follows: At the top of the supply chain, trust in a
specific version of a piece of software is established through auditing
the source code or, more likely, by implicitly trusting its developers
(e.g.~trusting version 5.11.5 of the Linux kernel as it is signed by Linus
Torvalds). Later, but before executing any binaries they have downloaded, users
can compare the checksums of these files with the expected values for that
specific version, crucially aborting on any mismatch.

These expected checksums could originate from a limited set of trusted parties
who publish statements that building some specific source code release results
in a particular set of executables. However, another alternative is
\emph{distributed consensus}, where a loose-knit community of semi-trusted
builders independently announce their checksums. Normally, participants in this
scheme report identical checksums for a given source code release, but in case
of a discrepancy (i.e.~if some builders have been compromised), the checksum
reported by $\geq 50\%$ of the builders may be the one to trust. Under this
verification scheme, the takeover at least 50\% of the builder community would
be required to co-erce users into running malicious binaries.

 \section{REPRODUCIBILITY IN THE SMALL}
\label{sec:bestiary}

How hard is it to ensure that independent builds always result in bit-for-bit
identical executables? First, let us consider the software vendor in
Figure~\ref{fig:approach}. Each build takes as its input the source to be
built, all of its build-time dependencies, and the entire build toolchain
including the compiler, linker and build system. The build produces a set
of artifacts (executables, data, documentation, etc.) as its output. Any change
in these inputs may legitimately affect its output.

However, even once all inputs have been controlled for, the build may still be
unreproducible---that is, producing different artifacts when the build is
repeated. This results from two main classes of problem: uncontrolled build inputs
and build non-determinism.

\emph{Uncontrolled build inputs} occur when toolchains allow the build process
to be affected by the surrounding environment. Common examples include system
time, environment variables and the arbitrary build location on the filesystem.
Uncontrolled inputs can be seen as analogous to breaking encapsulation in
software design; a tight coupling between a high-level process and the
low-level implementation details.

\emph{Build non-determinism} occurs when aspects of the build behave
non-deterministically \emph{and} these ``random'' behaviours are encoded in the
final artifacts. For example, if the output is derived in any way from the
state of a pseudorandom number generator or the arbitrary order of process
scheduling.

To address uncontrolled build inputs, it is tempting to ``jail'' builds into
sanitized environments that always present a canonical interface to the
underlying build system. Indeed, this was the approach taken by early projects
such as Bitcoin and Tor (\URL{rbm.torproject.org}). However, jails result in
slower build times and impose technical and social restrictions on developers
who may be accustomed to choosing their tooling. Most jails cannot address
non-determinism issues either.

The ultimate and preferred solution is to ensure that any code run during the
build \emph{only} depends on the legitimate build inputs (the source being
built, the build dependencies and the toolchain), and that any
non-deterministic behavior does not affect the resulting artifacts.

\medskip

We will now review some individual causes of unreproducible builds and show how
to address them.

\paragraph{Build timestamps}

\def\captionExDate{The \texttt{\_\_DATE\_\_} C preprocessor macro
  \emph{``expands to a string constant that describes the date on which the
    preprocessor is being run.''}}
\begin{lstlisting}[language=C,float,label=ex:date,caption=\captionExDate]

void usage() {
    fprintf (stderr,
             "foo-utils version "
             "3.141 (built %s)\n",
              __DATE__);
}
\end{lstlisting}

Timestamps are, by far, the biggest source of unreproducibility. It is a common
practice to explicitly embed dates into binaries via C's \texttt{\_\_DATE\_\_}
macro (see Listing~\ref{ex:date}), but many tools record dates into build
artifacts as well. For example, \texttt{help2man} generates UNIX manual pages
directly from the output of \texttt{--help}, and in its default configuration,
it embeds the current date into generated files. As this value changes from
day-to-day, this results in an unreproducible build. \TeX{}'s
\texttt{\textbackslash date} macro also embeds the current date, with similar
implications for generated documentation.

The value of these timestamps is extremely limited, particularly as they don't
convey which version of the software was actually built; after all, older
software can always be built later. Embedded timestamps should therefore be
avoided entirely, but for cases where that is not possible (e.g.~in file
formats that mandate their presence), the Reproducible Builds project proposed
the \texttt{SOURCE\_DATE\_EPOCH} environment variable as a way to communicate
an acceptable timestamp to build systems~\cite{source-date-epoch}. This
typically represents the last modification time of the source tree as extracted
from the software's changelog file.

\paragraph{Build paths}

The filesystem path where the build took place is often embedded in generated
binaries too, usually via the \texttt{\_\_FILE\_\_} preprocessor macro (see
Listing~\ref{ex:file}) or by assertion statements that reference their
corresponding line of code. Other sources of build paths include logging
messages, locations of detached debug symbols, \texttt{RPATH} entries in ELF
binaries, and many other instances that are intended, ironically, to assist the
software development process.

\def\captionExFile{The \texttt{\_\_FILE\_\_} C preprocessor macro
  \emph{``expands to the name of the current input file''}. This results in
  non reproducibility when the program is built from different directories,
  e.g.~\texttt{/home/lamby/tmp} vs.~\texttt{/home/zack/tmp}.}
\begin{lstlisting}[language=C,float,label=ex:file,caption=\captionExFile]

fprintf (stderr,
         "DEBUG: boop (%s:%s\n",
         __FILE__, __LINE__);
\end{lstlisting}

To help address this issue, the Reproducible Builds project worked with the
GNU GCC developers to introduce the \texttt{-ffile-prefix-map} and
\texttt{-fdebug-prefix-map} options which support embedding relative (rather than
absolute) paths.

\paragraph{Filesystem ordering}

Contrary to the output of \texttt{ls(1)}, the POSIX Unix standard does not
specify an ordering for results returned by the underlying \texttt{readdir(3)}
system call. As a result, directories accessed in naive ``\texttt{readdir}
order'' may be processed in a non-deterministic manner. If this arbitrary
ordering influences any build artifacts, the build will not be reproducible.

For example, the build system of the \emph{PikePDF} library located its own
source files using Python's \texttt{glob} routine. But as \texttt{glob}'s
result value inherits the non-determinism of \texttt{readdir(3)},
\emph{PikePDF}'s source files were linked in an arbitrary order.

This is a particularly pernicious problem as some filesystem implementations
return different orderings ``more often'' than others. To avoid these issues,
build systems should impose a deterministic order on any directory iteration
encoded in its artifacts, e.g.~via an explicit \texttt{sort()}.

\paragraph{Archive metadata}

\texttt{.zip} and \texttt{.tar} archives store timestamps and user ownership
information in addition to the files themselves. However, if this metadata is
inherited from the surrounding build environment, it will not be replicated
when building elsewhere. For example, if a \texttt{.tar} archive stores files
as belonging to the build user (e.g.~\texttt{lamby}), another user
(e.g.~\texttt{zack}) building the same software will obtain a different result.

This can be avoided by instructing tools to ignore on-disk values in favour
of metadata chosen by the build system (e.g.~using \texttt{tar(1)} with
\texttt{--owner=0} and \texttt{--clamp-mtime=T}), or by normalizing metadata
before archiving begins (e.g.~by using \texttt{touch(1)} with
\texttt{SOURCE\_DATE\_EPOCH} as a reference timestamp).

\paragraph{Randomness}

\def\captionExHash{
   Perl's hash type does not define an ordering of its keys, so a call to
  \texttt{sort} should be inserted before \texttt{keys \%h} to make it
  deterministic.}
\begin{lstlisting}[language=Perl,float,label=ex:hash,caption=\captionExHash]

my %h = ( a => 1, b => 2, c => 3);
foreach my $k (keys %h) {
    print "$k\n";
}
\end{lstlisting}

Even when the entire environment is controlled for, many builds remain
inherently non-determinstic. For example, builds that iterate over hash tables
(such as Perl's ``hash'' or the \texttt{dict} type in Python $<$\,3.7) exhibit
arbitrary behaviour as their respective elements are returned in an undefined
order---the code in Listing~\ref{ex:hash}, for example, may print any
combination of \texttt{abc}, \texttt{bac}, \texttt{bca}, etc. This affects
reproducibility if these results form any part of the build's artifacts.

Parallelism (such as via processes or threads) can also prevent reproducibility
if the arbitrary completion order is encoded into build results too. Similar to
filesystem ordering, these issues can be resolved by imposing determinism
in key locations, seeding any sources of randomness to fixed values or
sorting the results of hash iterations and parallelized tasks before generating
output.

\paragraph{Uninitialized memory}

Many data structures have undefined areas that do not affect their operation.
The FAT filesystem, for example, contains unused regions that may be filled with
arbitrary data. In addition, modern CPU architectures perform more efficiently
when data is naturally aligned, and the padding added to ensure alignment can
result in similarly undefined areas. These regions containing ``random'' data affect
reproducibility when stored in build results.

\def\captionExMemset{A patch for \emph{GNU mtools} ensuring that a
  \texttt{direntry\_t} struct does not contain uninitialized memory.}
\begin{lstlisting}[language=C,float,label=ex:memset,caption=\captionExMemset]

--- a/direntry.c
+++ b/direntry.c
@@ -24,6 +24,7 @@
 
 void initializeDirentry(
   direntry_t *entry, Stream_t *Dir) {
+  memset(entry, 0, sizeof(direntry_t));
   entry->entry = -1;
   entry->Dir = Dir;
\end{lstlisting}

One solution is to explicitly zero-out memory regions that may persist in
artifacts. For example, Listing~\ref{ex:memset} shows a patch for \emph{GNU
  mtools} that ensures generated FAT directory entries do not embed
uninitialized memory.

 \section{REPRODUCIBILITY IN THE LARGE}
\label{sec:qa}

\def\captionExHash{An example \BUILDINFO file, recording both the environment and
  results of building Debian's \texttt{black} package. (Excerpt: see
  \URL{buildinfo.debian.net/sources/black/20.8b1-1} for the full version.)}
\begin{lstlisting}[float=*,escapechar=|,label=ex:buildinfo,caption=\captionExHash]

|\textbf{\texttt{Source}}|: black
|\textbf{\texttt{Version}}|: 20.8b1-1
|\textbf{\texttt{Checksums-Sha1}}|:
  9915459ae7a1a5c3efb984d7e5472f7976e996b1 2584 black_20.8b1-1.dsc
  14bfd3011b795f85edbc8cc4dc034a91cfaa9bcd 111096 black_20.8b1-1_all.deb
  69c3d4ae7115c51e7b00befe8b4afd5963601d66 285684 python-black-doc_20.8b1-1_all.deb
|\textbf{\texttt{Checksums-Sha256}}|: [...]
|\textbf{\texttt{Build-Architecture}}|: amd64
|\textbf{\texttt{Installed-Build-Depends}}|: autoconf (= 2.69-11.1), automake (= 1:1.16.2-4), [...],
  gcc (= 4:10.2.0-1), [...], python3 (= 3.8.2-3), [...]
  xz-utils (= 5.2.4-1+b1), zlib1g (= 1:1.2.11.dfsg-2)
\end{lstlisting}

Now that we know how to address some individual reproducibility issues, we turn
to the problems that arise when making large software \emph{collections}
reproducible.

The Reproducible Builds project started in 2014 with the aim of making the
Debian operating system (\URL{www.debian.org}) completely reproducible. This is
a formidable goal, as not only is Debian a extremely mature Linux distribution,
it is one of the largest curated collections of FOSS software in general.

Seven years later, over 95\% of the \num{30000}+ packages in Debian's
development branch can now be built reproducibly, and as the Linux distribution
with the largest total number of reproducible packages, it serves as an
extremely relevant case study. The evolution of this effort can be found at
\URL{wiki.debian.org/ReproducibleBuilds}.

\paragraph{Adversarial rebuilding}

Given its scale, Debian developers realized they would need a programmatic way
to test for reproducibility. To this end, they developed a continuous integration
(CI)~\cite{meyer2014ci} system which builds each package in the Debian archive
twice in a row, using two independent build environments that are deliberately
configured to differ as much as possible. For instance, the clock on the second
build is set 18 months in the future, and the hostname, language, system
kernel, etc., are all varied so that if any environmental differences are used
as a build input, the two builds will differ as a result.  The large number of
variations applied (30+) can validate build reproducibility to a high degree of
accuracy.

To identify any reliance on non-deterministic filesystem ordering, the R-B
project also developed a FUSE-based~\cite{vangoor2017fuseperf} virtual
filesystem called \emph{disorderfs}
(\URL{salsa.debian.org/reproducible-builds/disorderfs}) which can provide a
view of a filesystem with configurable orderings. The R-B CI system
\emph{reverses} the filesystem ordering between the builds, revealing any
dependency on non-deterministic filesystem ordering.

\paragraph{Recording build information}

As per Definition~\ref{def:r-b}, a reproducible build must always use the same
original source, toolchain and build dependencies, and to ensure these inputs
can be replicated correctly, Debian devised the \BUILDINFO file format.

Once a Debian package is built, the precise source version and the versions of
all its build dependencies are recorded in a \BUILDINFO file. This file also
contain checksums of any generated \texttt{.deb} artifacts, the Debian binary
package format. (An example file may be found in Listing~\ref{ex:buildinfo}.)

\BUILDINFO files are a crucial building block for any process wishing to
validate reproducibility. A \BUILDINFO is produced during an initial build and
is then used to reconstruct a second build environment. The build is repeated
within this second environment and the checksums from this latter build are
compared with the ones in the \emph{original} \BUILDINFO---if they do not
match, the build is unreproducible or a build host has been tampered with.

\begin{figure*}
  \centering
  \includegraphics[width=\textwidth]{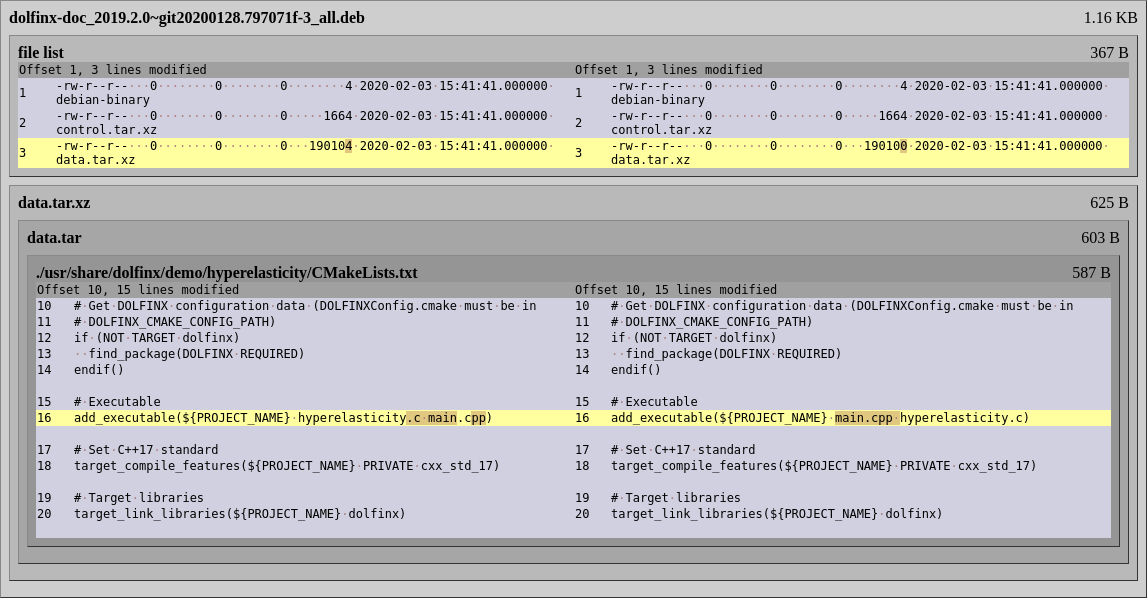}
  \caption{\emph{diffoscope} recursively unpacks archives of many kinds and
    transforms various binary formats into more human-readable forms in order
    to compare them.}
  \label{fig:diffoscope}
\end{figure*}

Users can employ \BUILDINFO files to implement the consensus-driven approach
outlined above, verifying downloaded packages by comparing them against the
checksums in \BUILDINFO files distributed by Debian and other builders. In this
scenario, \BUILDINFO files are cryptographically signed to represent a
\emph{build attestation}, e.g.~``I, Alice, given source X and environment Y,
have built a package with checksum K.'' Bob would verify that Alice really made this
claim, and then compare Alice's \textit{K} against his downloaded file,
potentially trusting Alice's \textit{K} over any divergent (and likely
malicious) claim from Eve.

Debian currently hosts over 20 million \BUILDINFO files in a number of
experimental services. However, centralized distribution schemes inherit many
of the issues of the SSL certificate authority ecosystem, particularly in
representing an obvious target to attack~\cite{laurie2014ssl}. Decentralized
alternatives remain a future challenge at this point, as does a practical
consensus mechanism to determine the ``valid'' checksum for any given package.

\paragraph{Root cause analysis}

As we have outlined, it is trivial to detect mismatches between builds simply
by comparing the checksums of their artifacts. However, it can be extremely
difficult to understand the root cause of this difference.

Therefore, the R-B project developed \emph{diffoscope} (\URL{diffoscope.org}),
a visual ``diff'' tool that recursively unpacks a large number of archive
formats and translate tens of binary formats into human-readable forms. As a
result, it can display the meaningful, code-level differences between, for
example, two compiled Java \texttt{.class} files, even if they were contained
in \texttt{.tar} in a \texttt{.xz} (in a \texttt{.deb} in a
\texttt{.iso},~etc.).

In most cases, \emph{diffoscope} indicates which category of fix is required to
make a build reproducible. For instance, when programs embed build dates into
their binaries, \emph{diffoscope} clearly highlights these date-based
variations, and the surrounding context tends to assist in identifying which
part of the original source code to fix.

An example \emph{diffoscope} output is shown in Figure~\ref{fig:diffoscope},
where two versions of the \texttt{dolfinx-doc} package differ. Here,
\emph{diffoscope} indicates that the difference is in \texttt{CMakeLists.txt},
a generated file which contains the same entries with a different ordering
between the two builds. This would appear to be a filesystem ordering issue,
solved by the addition of an explicit sort.  However, this may be a problem
affecting all software that uses CMake, so the issue may be better addressed
there; alas, \emph{diffoscope} cannot entirely replace a software engineer's
judgement.

\paragraph{Quality Assurance (QA)}

Adopting a methodical approach to verify the reproducibility of builds can be
highly complementary to QA efforts. This is because problems that affect build
reproducibility are often symptoms of larger, systemic issues.

To begin with, systematic \emph{reproducibility} testing implies systematic
\emph{build} testing, so will easily identify software that fail to build under
any circumstances.  Other software will fail to build only in the extreme
environments designed to test for reproducibility, but will become more robust
as a result of behaving well there. For example, some software will fail to
build in the future due to hardcoded SSL certificates with expiry dates---these
are detected due to the build environment's artificial future clock. Others
fail to build in rarely-used timezones due to incorrect assumptions about time
offsets---the test suite for the Ruby \emph{Timecop} library failed in this way
(\DEBBUG{795663}).

Due to these serendipitous quality improvements, addressing reproducibility
issues improves the correctness and robustness of the Debian distribution as a
whole. However, even less-critical problems can be identified through
reproducibility testing as well. For example, an issue in the Doxygen
documentation generator (\DEBBUG{970431}) led to broken hyperlinks that linked
to their build-time location (e.g. \texttt{/tmp/build/foo/usage.html}) instead
of their run-time one (\texttt{/usr/share/doc/foo/usage.html}). This was
trivial to identify with \emph{diffoscope}, and fixing it corrected broken
documentation for hundreds of end-users.

Reproducibility testing can even flag spurious content \emph{within}
documentation. For instance, manual packages generated by executing an underlying
program can fail in several ways, often printing error messages that are
mistakenly shipped as the package's ``documentation'' (e.g.~\DEBBUG{972635}).
These bugs are difficult to detect if the failure does not occur on a
developer's own machine, but they can be easily spotted whilst testing for
reproducibility as the error messages are deliberately designed to appear in
different languages.

\def\captionExGbrowse{An example \texttt{ConfigData.pm}. As it was created at
 build time, all users shared the same \texttt{OpenIDConsumerSecret}.}
\begin{lstlisting}[float,label=ex:gbrowse,caption=\captionExGbrowse,basicstyle=\footnotesize\ttfamily]

{
 'cgibin' => '/usr/lib/cgi-bin/gbrowse',
 'conf' => '/etc/gbrowse',
 'databases' => '/var/lib/gbrowse/databases',
 'htdocs' => '/usr/share/gbrowse/htdocs',
 'OpenIDConsumerSecret' => '639098210478536',
 'tmp' => '/var/cache/gbrowse'
},
\end{lstlisting}

Even security issues can be discovered whilst testing for reproducibility. In
one example, the \emph{GBrowse} biological genome annotation viewer failed to
build reproducibly (\DEBBUG{833885}), and \emph{diffoscope} identified a
configuration file that contained a different \texttt{OpenIDConsumerSecret}
value between builds (see Listing~\ref{ex:gbrowse}). Although this secret was
being securely generated, it was being created at \emph{build time}, so the
same value was distributed to all users of the package---the fix was to
generate the secret at installation time so that each deployment possessed its
own unique key. The mechanics of reproducibility testing suggest that this
issue would not have been readily discovered another way.

\paragraph{Community engagement}

Although the causes of build unreproducibility often reside within the source
code of individual projects, it is far more effective to detect issues via
centralized testing in distributions such as Debian due to the uniform build
interfaces these large collections provide. Nevertheless, the social norms of
the FOSS community dictate that fixes should be integrated upstream, instead of
remaining in distribution-specific patch sets.

To this end, the Reproducible Builds project has contributed to hundreds of
individual FOSS projects, in addition to working with key toolchains such as
GCC, Rust, OCaml,~etc. This community-oriented approach ensures that as many
users as possible can benefit from the specific advantages of reproducible
builds, as well as from the software quality improvements achieved while pursuing that
goal.

 \section{THE REPRODUCIBLE BUILDS ECOSYSTEM}

Taking Debian to its current state required over seven years of cross-community
work that was spearheaded by the Reproducible Builds project
(\URL{reproducible-builds.org}), a non-profit organisation that aims to
increase the integrity of software supply chains by advocating for and
implementing the approach outlined in this paper.

Although originating in Debian around 2014, many other FOSS projects have
joined the initiative such as Arch Linux, coreboot, F-Droid, Fedora, FreeBSD,
Guix, NixOS, openSUSE and Qubes. One milestone of this joint effort is Tails
(\URL{https://tails.boum.org/}), the operating system used by Edward Snowden to
securely communicate the NSA's global surveillance activities in 2013~\cite{landau2013snowden}---Tails
began releasing reproducible ISO images in 2017 to improve end-user
verifiability and security.

The Reproducible Builds project has also developed several tools
(\URL{reproducible-builds.org/tools}) that facilitate various QA processes
related to reproducibility. Some of these, such as \emph{diffoscope} and
\emph{disorderfs}, have been highlighted in this paper.

Increasing the security of open source software is clearly a worthwhile goal,
and software professionals and organisations can always provide assistance.
This is not only by addressing any uncontrolled build inputs and sources of
non-determinism in the software they maintain, but by working with the
Reproducible Builds project itself in terms of code, donations and other
traditional forms of community contribution.

\section{CONCLUSION}
\label{sec:conclusion}

In this article, we have outlined what it means for software to build
reproducibly and how that property can be leveraged by end-users to establish
trust in open source executables, even when they are built by untrusted third
parties. We also surveyed several causes of unreproducibility and located their
causes in build systems and similar logic. We also described some of the
quality assurance (QA) processes and tools that can be used to make large open
source software collections reproducible---using this model, the Debian
operating system has achieved 95\% reproducibility in over \num{30000}+
packages.

Additional work is still needed to address the software that is not yet
reproducible. In the case of Debian, there are no insurmountable obstacles
preventing the project from reaching 100\%---the remaining 5\% ``only'' need
fixes similar in kind to those already discussed. However, this has not yet
been achieved, partly because time and effort are not inexhaustible or fungible
resources in volunteer communities, but also due to regressions in
previously-reproducible packages. Improved awareness and prioritisation of
reproducibility amongst software developers would reduce the incidence of such
events.

Other challenges remain for the reproducible builds ecosystem too.
Cryptographically signed artifacts are becoming more common, which cannot be
made reproducible without distributing signing keys to builders. One solution
is to adopt detached signatures, but the addition of parallel distribution
channels for these (unreproducible) files would require extensive changes to
existing software distribution channels.

The verification of open source software for mobile devices also remains
problematic. With the notable exception of F-Droid, not only are the build
processes of the major app stores unreproducible (or not even FOSS), the
checksums of artifacts are hidden from end-users, rendering any distributed
validation scheme impossible. Significant usability and transparency
improvements are needed to make meaningful progress in this area.

Finally, we are left with the recursive question of whether we can trust
even \emph{reproducible} binaries without trusting where our compilers and
other toolchain components come from. To address this, the parallel
Bootstrappable Builds (\URL{bootstrappable.org}) project seeks to minimize the
amount of binary code required to bootstrap a minimal C compiler---at time of
publication, a binary as small as 6\,KB is enough to activate a chain of steps
from \emph{TCC}~\cite{bellard2003tcc} to GCC from which almost all toolchains
can then be obtained. Ken Thompson would likely approve, whilst still pointing
out that 6\,KB is too much untrusted code.

\section{ACKNOWLEDGMENTS}
\label{sec:acks}

The authors would like to thank the Reproducible Builds and the wider Debian
community for their feedback on this paper, as well as for their invaluable
work on increasing the trustworthiness of free and open source software. The
authors also thank Giovanni Mascellani for their insightful discussions on
bootstrappable builds.

\begin{IEEEbiography}{Chris Lamb}{\,} is a freelance programmer with over
  fifteen of experience of developing open source software. He has contributed
  to the Debian operating system since 2006 and was elected to serve as the
  Project Leader in 2017 and 2018. He is also a director of the Open Source
  Initiative (OSI) as well as Software in the Public Interest (SPI), Inc.
  Today, he is now highly active in the Reproducible Builds project, through
  which he has received a grant from the Linux Foundation. Contact him at
  chris@chris-lamb.co.uk.
\end{IEEEbiography}

\begin{IEEEbiography}{Stefano Zacchiroli}{\,} is Associate Professor of
  Computer Science at Université de Paris on leave at Inria, France. He is
  co-founder and current CTO of the Software Heritage project. He is a member
  of the steering committee of the Reproducible Builds project. He has served
  as Debian Project Leader over the period 2010-2013. Contact him at
  zack@irif.fr.
\end{IEEEbiography}

\end{document}